\documentclass[iop]{emulateapj}

\usepackage{relsize}
\usepackage{amsmath}
\usepackage{hyperref}

\shorttitle{Linking the \textsc{x3d} pathway to integral field spectrographs}
\shortauthors{Vogt et al.}

\begin{document}

\submitted{Accepted for publication in PASP's ``Techniques and Methods for Astrophysical Data Visualization'' special issue}

\title{Linking the \textsc{x3d} pathway to integral field spectrographs: \\
    YSNR 1E\,0102.2-7219 in the SMC as a case study}

\author{Fr\'ed\'eric P.A. Vogt\altaffilmark{$\star$, $\dagger$, 1} and Ivo R. Seitenzahl\altaffilmark{2,3} and 
Michael A. Dopita\altaffilmark{2} and Ashley J. Ruiter\altaffilmark{2,3}}

\altaffiltext{$\star$}{Contact: frederic.vogt@alumni.anu.edu.au}
\altaffiltext{$\dagger$}{ESO Fellow}
\altaffiltext{1}{European Southern Observatory, Av. Alonso de C\'ordova 3107, 763 0355 Vitacura, Santiago, Chile. }
\altaffiltext{2}{Research School of Astronomy and Astrophysics, Australian National University, Canberra, Australia.}
\altaffiltext{3}{ARC Centre of Excellence for All-sky Astrophysics (CAASTRO).}

\begin{abstract}
The concept of the \textsc{x3d} pathway was introduced by \cite{Vogt2016} as a new approach to sharing and publishing 3-D structures interactively in online scientific journals. The core characteristics of the \textsc{x3d} pathway are that: 1) it does not rely on specific software, but rather a file format (\textsc{x3d}), 2) it can be implemented using fully open-source tools, and 3) article readers can access the interactive models using most main stream web browsers without the need for any additional plugins. In this article, we further demonstrate the potential of the \textsc{x3d} pathway to visualize datasets from optical integral field spectrographs. We use recent observations of the oxygen-rich young supernova remnant 1E\,0102.2-7219 in the Small Magellanic Cloud to implement additional \textsc{x3dom} tools \& techniques and expand the range of interactions that can be offered to article readers. In particular, we present a set of \textsc{javascript} functions allowing the creation and interactive handling of clip planes, effectively allowing users to take measurements of distances and angles directly from the interactive model itself.  
\end{abstract}

\keywords{ISM: supernova remnants; ISM: individual objects (SNR 1E\,0102.2-7219); techniques: miscellaneous }

\section{Introduction}

Oxygen-rich (O-rich) young supernova remnants (YSNRs) form a special subclass of supernova remnants (SNRs). In those systems, the observed forbidden-line oxygen emission is understood to arise from ejecta encountering (at several 1000\,km s$^{-1}$) the reverse shock wave from a type Ib supernova that had been stripped of its hydrogen envelope prior to the explosion \citep[][]{Sutherland1995}. O-rich YSNRs are as such unlike most SNRs where the optical emission comes from the interstellar medium ionized by the forward shock. Only a handful of such O-rich YSNRs are known: these include Cas A \citep{Chevalier1979}, G292.0+1.8 \citep{Goss1979,Murdin1979} and Puppis A \citep{Winkler1985} in our Galaxy, N132D \citep{Danziger1976,Danziger1976a,Lasker1978} \& SNR 0540-69.3 \citep{Mathewson1980} in the Large Magellanic Cloud, and 0103-72.6 \citep{Park2003}, B0049-73.6 \citep{Hendrick2005,Schenck2014} \& 1E\,0102.2-7219 \citep[1E\,0102 for short;][]{Dopita1981,Tuohy1983} in the Small Magellanic Cloud (SMC).

The ages of O-rich YSNRs are typically only of a few thousands of years, so that the O-bright ejecta have not (yet) had time to strongly interact with the surrounding medium. It is thus reasonable to assume that the ejecta have been \emph{freely expanding} since the time of the SN explosion \citep[see e.g.][]{Milisavljevic2013}. This property can be used to derive the ages of YSNRs through the measurement of their ejecta's proper motion, which display typical velocities of the order of several $1000$ km\,s$^{-1}$ \citep[][]{vandenBergh1970,Thorstensen2001,Fesen2006,Finkelstein2006,Winkler2009}.

When observed with integral field spectrographs (IFS), the \emph{free expansion} assumption allows the reconstruction of the full three-dimensional (3-D) spatial distribution of the O-rich ejecta in space. Provided that the age of the YSNR is known (e.g. through proper motion measurements and/or from historical records), the ejecta radial velocities (measured through Doppler shifts) can be trivially transformed into physical distances along the line of sight $z$ using the relation:
\begin{equation}\label{eq:v}
z = v_{los}\Delta t
\end{equation}
with $\Delta t$ the time since the SN explosion and $v_{los}$ the line-of-sight velocity of the ejecta. Assuming that the distance to the YSNR is also known, angular separations on-sky can be transformed into physical distances, thus leading to the full 3-D localization of the ejecta in space. \cite{Vogt2010} and \cite{Vogt2011} applied this technique to YSNR 1E\,0102 and N132D, respectively. \cite{Milisavljevic2013} used a similar approach (using numerous long-slit observations) to reconstruct the 3-D map of the ejecta in Cas A.

In this article, we exploit new IFS observations of YSNR 1E\,0102 to link the concept of the \textsc{x3d} pathway to data products from integral field spectrographs. The \textsc{x3d} pathway was first implemented by \cite{Vogt2014} and formally defined by \cite{Vogt2016} as a new way to share and publish 3-D astrophysical models interactively. The \textsc{x3d} pathway revolves around the \textsc{x3d} file format and the associated \textsc{x3dom} (pronounced \emph{X-Freedom}) framework to share 3-D models on the World Wide Web. Here, we use 1E\,0102 to complement the examples of \cite{Vogt2016} that applied the \textsc{x3d} pathway to H\,\textsc{\smaller I} observations of a compact group of galaxies observed with the Very Large Array. The very nature of YSNR 1E\,0102, and in particular the ability to reconstruct the 3-D map of its O-rich ejecta, effectively makes it the ideal candidate for demonstrating the potential of the \textsc{x3d} pathway when coupled to optical IFS datasets. Another implementation of the \textsc{x3d} pathway applied to the MUSE IFS observation of an extragalactic target is to be presented by Bellhouse et al. (in prep).

After introducing our IFS observations of YSNR 1E\,0102 in Sec.~\ref{sec:obs}, we describe our interactive 3-D map of the O-rich ejecta in this system in Sec.~\ref{sec:model}, and introduce new \textsc{x3dom}-based tools to interact with 3-D models in Sec.~\ref{sec:tools}. We summarize our conclusions in Sec.~\ref{sec:summary}.

\section{Revisiting YSNR 1E\,0102 with WiFeS}\label{sec:obs}

YSNR 1E\,0102 was observed with the WiFeS IFS \citep{Dopita2007,Dopita2010} -- mounted on the 2.3m Advanced Technology Telescope (ATT) of the Australian National University at Siding Spring Observatory \citep[NSW, Australia;][]{Mathewson2013} -- in 2016, August as part of a series of observations of SNRs in both the SMC and LMC (P.I.: Seitenzahl). YSNR 1E\,0102 had already been observed with WiFeS in late 2009 (P.I.: Dopita): that dataset then led to the first 3-D reconstruction of the O-rich ejecta in this system \citep{Vogt2010}. The most recent WiFeS observations of YSNR 1E\,0102 improved on the 2009 observations as follows: 1) they were performed using the second generations of CCDs in the WiFeS instrument, 2) the resulting 4-fields WiFeS mosaic does not contain any gap, and 3) the data were reduced using the \textsc{pywifes} data reduction pipeline \citep{Childress2014a,Childress2014}. The observation details and scientific implication of this new dataset are to be described in Seitenzahl et al. (in prep.). For the purpose of this article, it is sufficient to say that the 2016 WiFeS mosaics of YSNR 1E\,0102 is composed of 4 overlapping WiFeS fields (each of 25$\times$38 square arcsec with $1\times1$ square arcsec spatial pixels or \emph{spaxels}) with a total size of $56\times48$ square arcsec. The dataset is comprised of two mosaics (the \emph{red} and the \emph{blue}): a structure inherent of the dual-beam design of the spectrograph.

\begin{figure}[htb!]
\centerline{\includegraphics[width=\hsize]{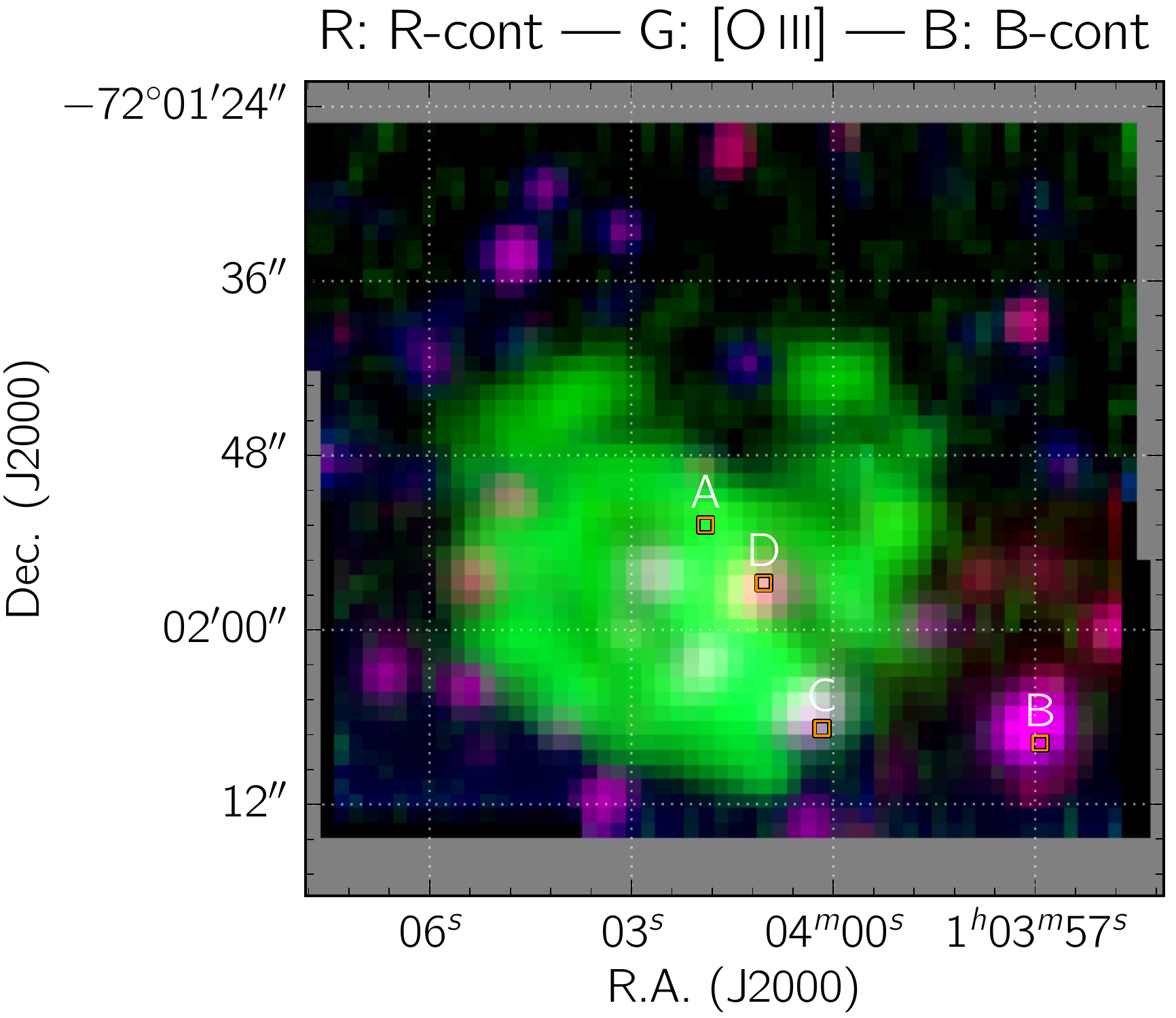}}
\caption{Pseudo-RGB image of 1E\,0102 reconstructed from the WiFeS mosaics. The R and B channels show the summed red and blue WiFeS mosaics respectively, while the G channel shows the integrated \& continuum subtracted [O\,\textsc{\smaller III}]$\lambda\lambda$4959,5007 emission (from which the SMC rest-frame emission was cropped). The orange rectangles and associated labels mark the location of four spaxels discussed in Fig.~\ref{fig:lowess} and \ref{fig:deblending}.}\label{fig:RGB}
\end{figure}

In this work, we solely focus on the blue mosaic unless explicitly mentioned otherwise. The observations were performed with the B7000 grating for the blue arm, resulting in a spectral resolution of R=7000  and a spectral coverage ranging from 4200\,\AA\ to 5548\,\AA. The seeing during the observations was of the order of 1.5\,arcsec. A pseudo-RGB image of YSNR 1E\,0102 highlighting the location of the O-rich ejecta in the system is presented in Fig.~\ref{fig:RGB}. The spatially extended emission from the fast moving ejecta of YSNR 1E\,0102 forms a complex set of structures (in green in the image). The numerous reddish-blue stars located across the field provide a visual indication of the spatial resolution of the data.

\subsection{A non-parametric continuum subtraction procedure}

\begin{figure*}[htb!]
\centerline{\includegraphics[scale=0.5]{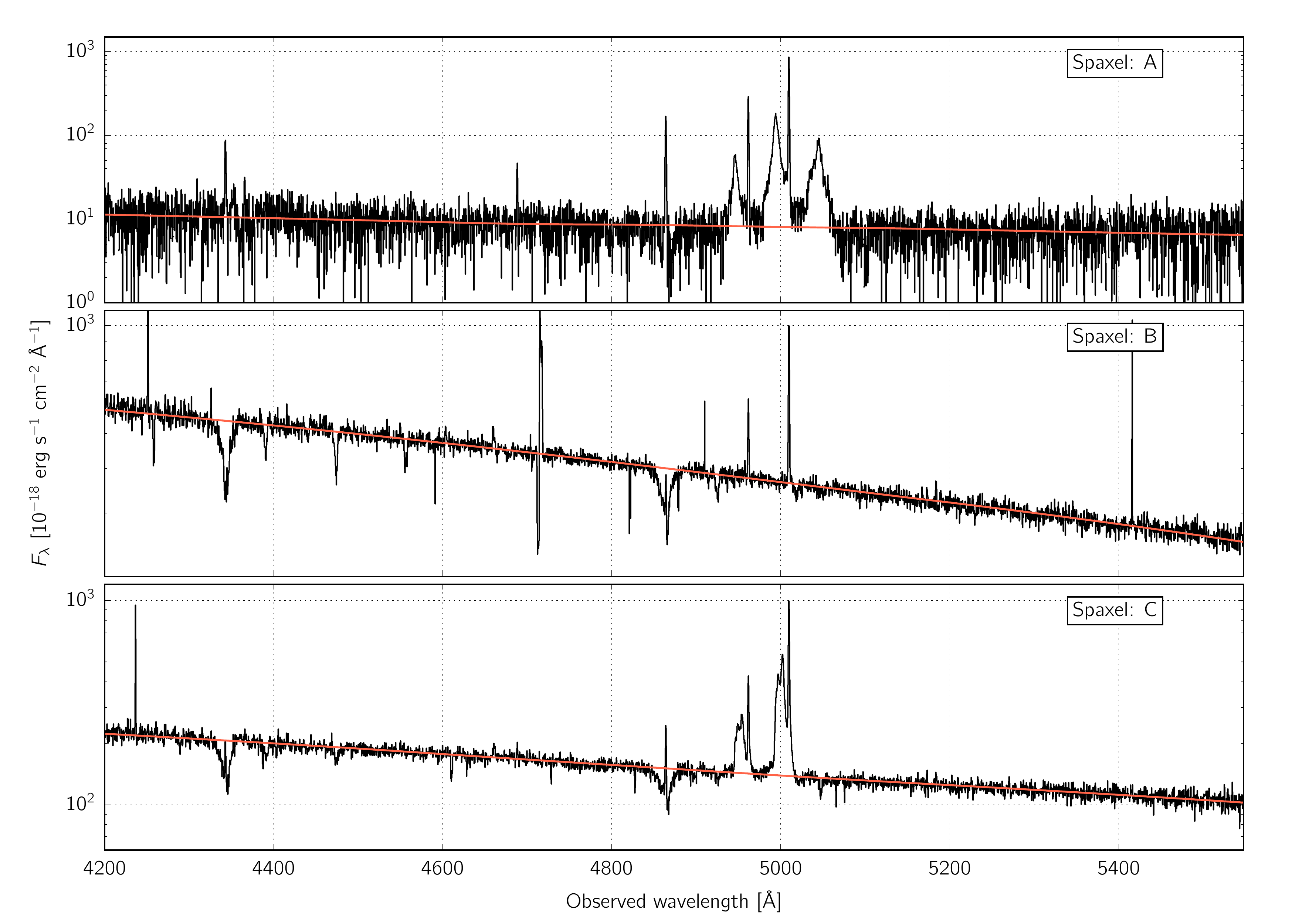}}
\caption{WiFeS spectra from the individual spaxels A, B and C (black curves; see Fig.~\ref{fig:RGB} for their location within the mosaic) and associated LOWESS fits to the continuum (red). The presence of strong \& broad [O\,\textsc{\smaller III}] emission and/or stellar absorption features does not affect the continuum fit, rendering this non-parametric technique particularly suitable to subtract the nebular and/or stellar continuum throughout the WiFeS mosaic on a spaxel-by-spaxel basis. }\label{fig:lowess}
\end{figure*}

Our approach for reconstructing the 3-D map of the O-rich ejecta in YSNR 1E\,0102 requires a continuum-subtracted datacube: both for spaxels dominated by nebular continuum, as well as for spaxels contaminated by bright stars. Three representative spectra extracted from individual spaxels and affected by bright O-rich ejecta and/or nebular continuum and/or bright stars are presented in Fig.~\ref{fig:lowess}. 

We rely on the locally weighted scatterplot smoothing (LOWESS) approach to fit the continuum for all spaxels in the blue mosaic (each spaxel being fitted individually). The LOWESS algorithm \citep[][]{Cleveland1979} is a non-parametric fitting technique that (through an iterative process) is only very weakly affected by bad pixels and/or strong emission \& absorption lines in a spectra. In practice, we use the \textsc{statsmodel} \citep[][]{Seabold2010} implementation of the LOWESS algorithm inside a custom-build \textsc{python} script. The resulting LOWESS fits are shown with red lines in Fig.~\ref{fig:lowess}. We stress that while the LOWESS fits do not remove stellar absorption features, the presence of these features in the continuum-subtracted datacube is of no importance for our subsequent analysis as they do not land within the [O\,\textsc{\smaller III}] spectral range.

\subsection{Deblending the [O\,\textsc{\smaller III}]$\lambda\lambda$4959,5007 lines}

Line-of-sight velocities $v_{los}$ for the O-rich ejecta in YSNR 1E\,0102 range from:
\begin{equation}
 -3500\text{\,km\,s$^{-1}$}<v_{los}<+5000\text{\,km\,s$^{-1}$}.
 \end{equation}
This implies Doppler shifts larger than 48\,\AA, the spectral gap between the [O\,\textsc{\smaller III}]$\lambda$4959 and [O\,\textsc{\smaller III}]$\lambda$5007 lines. As numerous sight-lines contain both blue- and redshifted material, the resulting total [O\,\textsc{\smaller III}]$\lambda\lambda$4959,5007 line profile is often complex (as illustrated in Fig.~\ref{fig:lowess}). As the intensity of the [O\,\textsc{\smaller III}]$\lambda$4959 and [O\,\textsc{\smaller III}]$\lambda$5007 forbidden emission lines are tied by a scaling factor of 2.98, their respective contributions to a given spectrum can nonetheless be disentangled through the following iterative process.

\begin{figure*}[htb!]
\centerline{\includegraphics[scale=0.5]{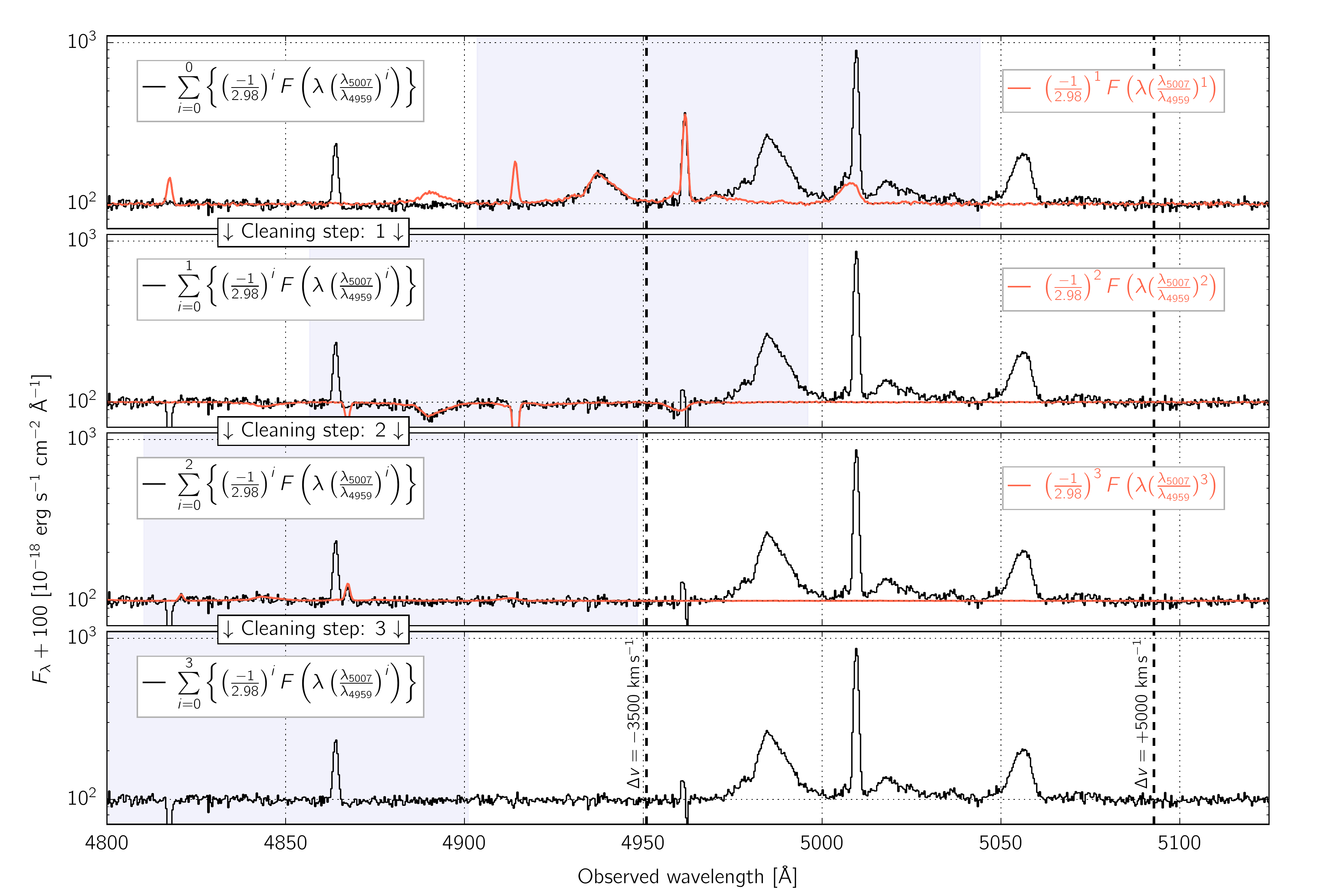}}
\caption{Demonstration of the iterative deblending procedure of the [O\,\textsc{\smaller III}]$\lambda\lambda$4959,5007 emission lines (for the first four steps). The black curves show the spectra of spaxel D (see Fig.~\ref{fig:RGB}) at cleaning step 0 to 4 (top to bottom). In each case, the red curve is the subsequent correction to be subtracted. Vertical dashed lines show the spectral range of the ejecta associated with the [O\,\textsc{\smaller III}]$\lambda$5007 line, spanning 8500\,km\,s$^{-1}$ from $\Delta v=-3500$\,km\,s$^{-1}$ to $\Delta v=+5000$\,km\,s$^{-1}$. The grey-colored range shows the spectral locations of the contamination associated with the [O\,\textsc{\smaller III}]$\lambda$4959 line. Only 2 cleaning steps are required for the [O\,\textsc{\smaller III}]$\lambda$5007 spectral features to be deblended from the [O\,\textsc{\smaller III}]$\lambda$4959 component. }\label{fig:deblending}
\end{figure*}

Let us define $F(\lambda)$ the continuum-subtracted flux density of a given sight-line as a function of the wavelength $\lambda$. Ignoring the presence of emission lines other than the [O\,\textsc{\smaller III}] lines, $F({\lambda})$ can be expressed as:
\begin{equation}\label{eq:Flam}
F(\lambda) = G\left(c\left(\frac{\lambda}{\lambda_{5007}}-1\right)\right)+\frac{1}{2.98}G\left(c\left(\frac{\lambda}{\lambda_{4959}}-1\right)\right)
\end{equation}
with $c$ the speed of light, and $\lambda_{5007}$ \& $\lambda_{4959}$ the rest wavelengths of the [O\,\textsc{\smaller III}]$\lambda$5007 \& [O\,\textsc{\smaller III}]$\lambda$4959 lines. $G(v_{los})$ represents the [O\,\textsc{\smaller III}]$\lambda$5007 flux density as a function of the ejecta velocity $v_{los}$: effectively, the disentangled [O\,\textsc{\smaller III}] spectra. Re-arranging Eq.~\ref{eq:Flam}, $G(v_{los})$ can be expressed as:
\begin{eqnarray}
G\left(v_{los}\right) &=& G\left(c\left(\frac{\lambda}{\lambda_{5007}}-1\right)\right)\nonumber\\
&=&\sum_{i=0}^{\infty}\left\{\left(\frac{-1}{2.98}\right)^iF\left(\lambda \left(\frac{\lambda_{5007}}{\lambda_{4959}}\right)^i\right)\right\} \label{eq:clean}
\end{eqnarray}
In words, disentangling the contributions from the [O\,\textsc{\smaller III}]$\lambda$4959 and [O\,\textsc{\smaller III}]$\lambda$5007 lines from a given spectrum requires to add \emph{shifted \& scaled} copies of the spectrum to the original one. Through this iterative process, the \emph{contamination} from the [O\,\textsc{\smaller III}]$\lambda$4959 line is shifted towards shorter wavelengths by a factor of $\lambda_{5007}/\lambda_{4959}$ and scaled by $2.98^{-1}$ at each step. This process is illustrated in Fig.~\ref{fig:deblending} for the first four steps of the summation in Eq.~\ref{eq:clean} for a representative WiFeS spectrum containing both red- and blueshifted ejecta. Given the range of radial velocities of the O-rich ejecta in YSNR 1E\,0102, only two cleaning steps ($i_{max}=2$) are required to disentangle the [O\,\textsc{\smaller III}]$\lambda$4959 and [O\,\textsc{\smaller III}]$\lambda$5007 lines for YSNR 1E\,0102. 

The presence of emission lines other than [O\,\textsc{\smaller III}] (e.g. H$\beta$) -- and for some spaxels of stellar absorption features -- leads to the creation of artefacts in the cleaned spectra: the clearest example is found at $\sim$4820\,\AA, and is caused by H$\beta$ during the first cleaning step. As these artefacts also move towards shorter wavelengths with each cleaning step, the lack of features red-wards of [O\,\textsc{\smaller III}]$\lambda$5007 ensures that the final disentangled [O\,\textsc{\smaller III}] line profile is \emph{clean}. 

\cite{Vogt2010} followed a similar disentangling procedure, but only performed one cleaning iteration. This was not strictly speaking sufficient to remove all contaminations from the [O\,\textsc{\smaller III}]$\lambda$4959 line (as illustrated in the second panel of Fig.~\ref{fig:deblending}) for all sight-lines. However, the remaining (negative) contamination where sparse and did not affect their analysis \& conclusions.

\section{Constructing the 3-D map of the O-rich ejecta}\label{sec:model}

Having fully removed the contamination from the  [O\,\textsc{\smaller III}]$\lambda$4959 line from each spectra, the WiFeS datacube can be directly converted in a 3-D map with units of [pc$\times$pc$\times$pc] via Eq.~\ref{eq:v}, assuming a distance to the SMC of 62\,kpc \citep[][]{Graczyk2014,Scowcroft2016} and an age of 2050\,yr for YSNR 1E\,0102 \citep[][]{Finkelstein2006}. We rely on the \textsc{mayavi} module in \textsc{python} \citep[][]{Ramachandran2011} to create an interactive diagram of the O-rich ejecta, and export it to the \textsc{x3d} file format. This forms the first step of the \textsc{x3d} pathway, as described by \cite{Vogt2016}. We refer the reader interested in using \textsc{mayavi} to implement the \textsc{x3d} pathway to the demonstration scripts\footnote{\href{http://dx.doi.org/10.5281/zenodo.45079}{DOI: 10.5281/zenodo.45079}} published by \cite{Vogt2016} and hosted on a dedicated Github repository\footnote{\url{http://fpavogt.github.io/x3d-pathway}}. 

We do not here describe our \textsc{python} script at any length, beside the fact that we rely on the \textsc{mlab.contour3d} routine inside \textsc{mayavi} for drawing the 3-D isocontours of the [O\,\textsc{\smaller III}]$\lambda$5007 line flux density. This approach does not involve any fitting of the [O\,\textsc{\smaller III}]$\lambda$5007 line profile: we directly convert the \textit{spectral} extent of the emission line profile into a \textit{spatial} extent. This is motivated by the fact that the internal velocity dispersion of individual clumps $\sigma_v\lesssim100$\,km\,s$^{-1}$ is smaller than the observed width of the spectral structures of the [O\,\textsc{\smaller III}]$\lambda$5007 emission line profiles (typically $\gtrsim500$\,km\,s$^{-1}$), i.e. WiFeS resolves real \textit{velocity} stretches (and thus \textit{spatial} stretches along the line-of-sight) of spatially unresolved clumps of ejecta.

We use \textsc{mlab.quiver3d} for drawing additional elements in the model that include directional arrows, large wireframe spheres of 6\,pc and 12\,pc in radius (intended as scale reference for the model), and a black sphere of 1\,pc in diameter marking the \emph{center} of the model. We note that both \textsc{mlab.contour3d} and \textsc{mlab.quiver3d} are well handled by the \textsc{x3d}-exporter\footnote{Based on the Visualization ToolKit X3D exporter v0.9.1} of \textsc{mayavi}, the current specific limitations of which were discussed in details in the online examples provided by \cite{Vogt2016}. 

The center of the model is defined (for the X$\equiv$East-West and Y$\equiv$North-South directions) as the ejecta's origin derived from their proper motions \citep[][]{Finkelstein2006}. The model reference along the Z$\equiv$line-of-sight direction is set at the SMC rest-frame, derived by fitting a Gaussian profile to the H$\beta$ line profile integrated across the entire WiFeS mosaic.

\begin{figure}[htb!]
\centerline{\includegraphics[width=\hsize]{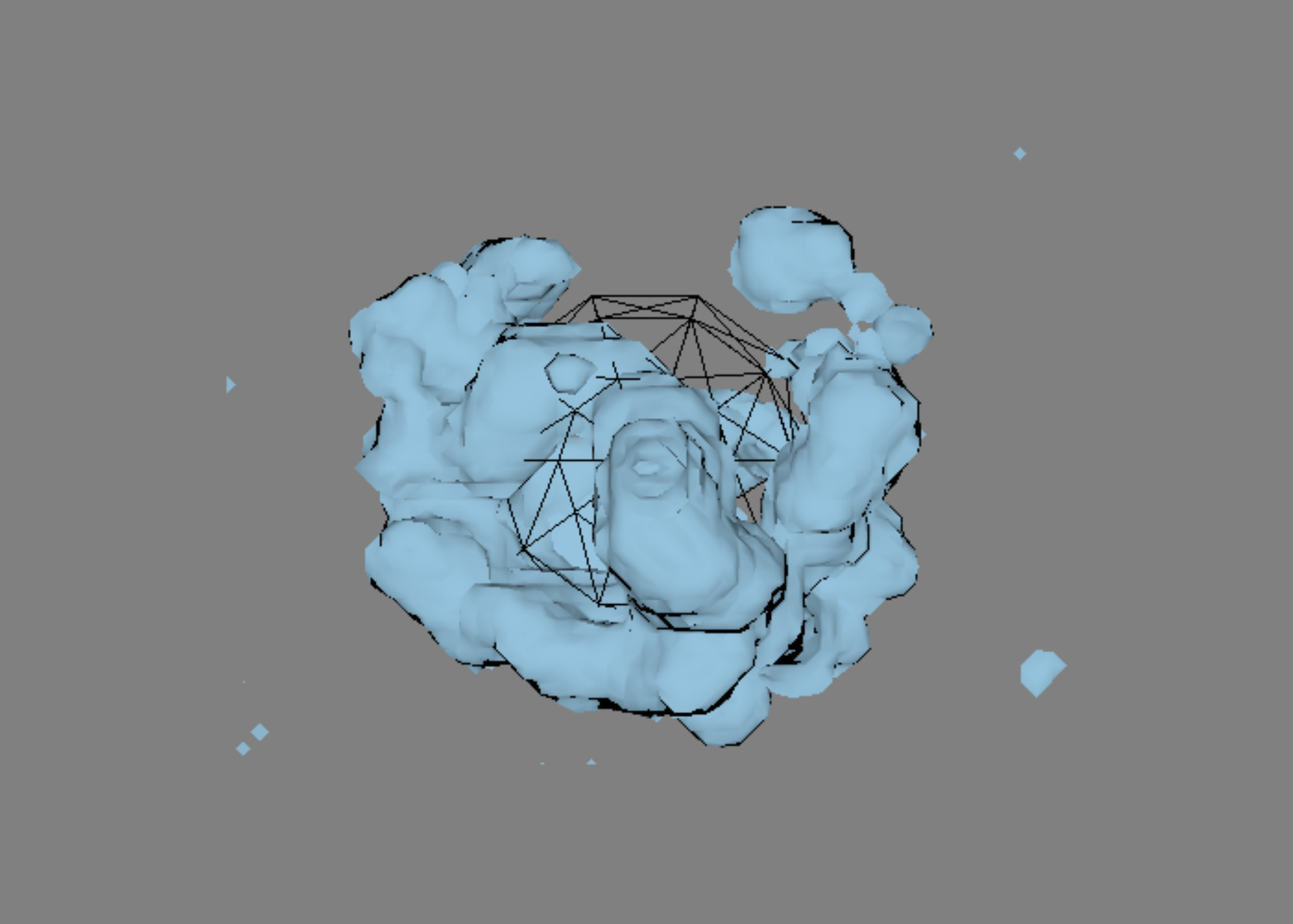}}
\centerline{\includegraphics[width=\hsize]{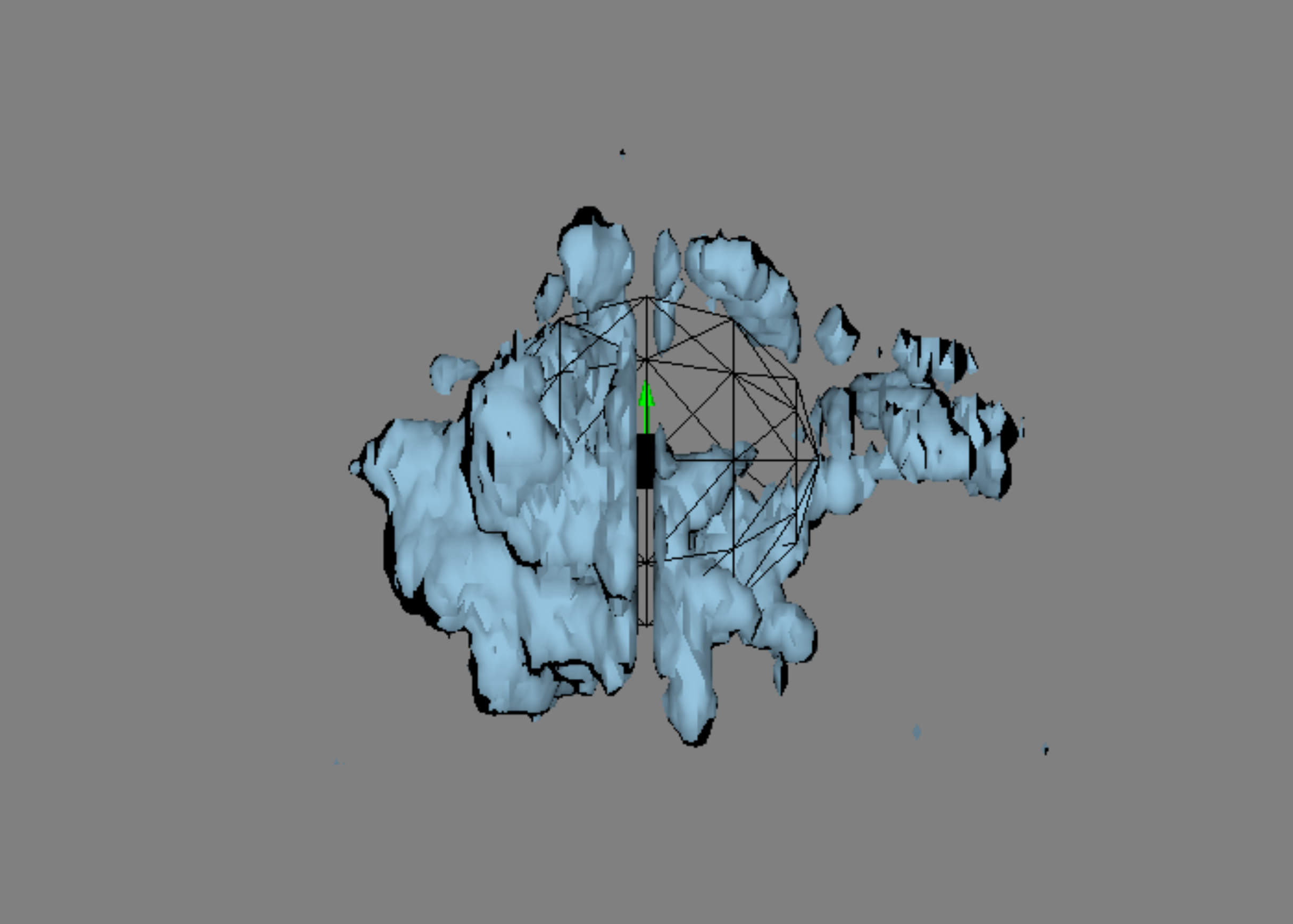}}
\centerline{\includegraphics[width=\hsize]{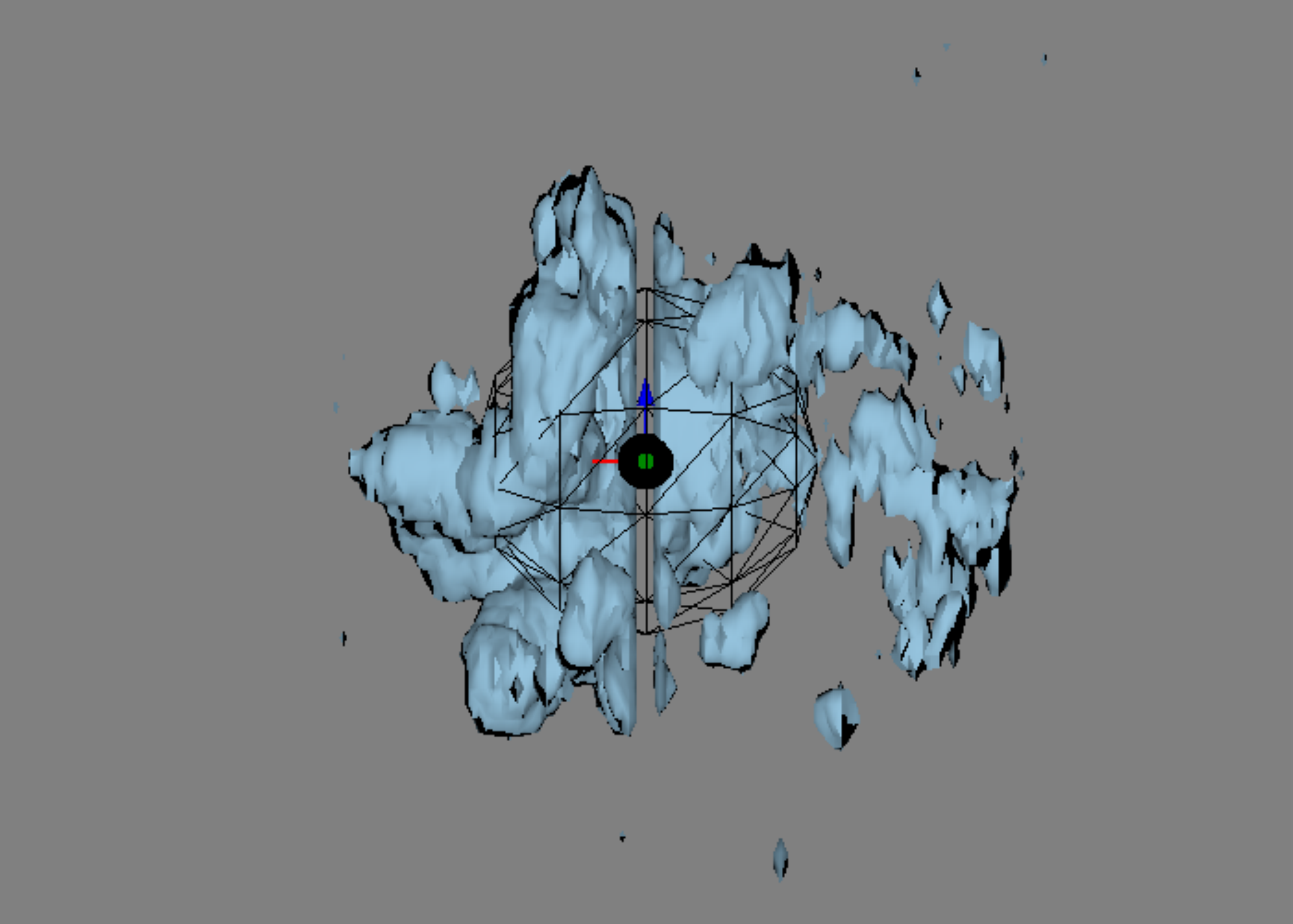}}
\caption{Reconstructed 3-D map of the [O\,\textsc{\smaller III}]$\lambda$5007 emission in 1E\,0102 at the level of $3.5\times10^{-17}$ erg\,s$^{-1}$\,cm$^{-2}$\,\AA$^{-1}$, as seen from the Earth (top, to be compared with Fig.~\ref{fig:RGB}), the West (middle) and the North (bottom). The central black sphere is 1\,pc in diameter, and marks the location of the supernova explosion derived from the ejecta proper motion (for the X-Y location), and set in the SMC rest-frame (for the Z direction). The larger wireframe sphere is 6\,pc in diameter, and included in the model as a scale reference. Most [O\,\textsc{\smaller III}]-bright ejecta are located outside of this 6\,pc sphere. In all panels, the red arrow points towards the Earth, the green arrow points North and the blue arrow points East. For demonstration purposes, these screenshots were generated using the \emph{Screenshot} tool on the \textsc{html} page hosting the interactive model. \textit{Note on the arXiv version: until the publication of the article, the online model can be accessed at \url{http://fpavogt.github.io/x3d-pathway/YSNR.html}}.}\label{fig:projections}
\end{figure}

Projections of the 3-D map (as seen from the Earth, North and West) are presented in Fig.~\ref{fig:projections}. There certainly exists a plethora of tools and techniques to create two-dimensional (2-D) projections of 3-D structures. For demonstration purposes, the illustrations in Fig.~\ref{fig:projections} were created using the \emph{screenshot} capabilities of the online, interactive version of the model. Until the publication of the article, the online model can be accessed at \url{http://fpavogt.github.io/x3d-pathway/YSNR.html}. 

Creating an online interactive \textsc{html} page (exploiting the \textsc{x3dom} framework) constitutes the second phase of the \textsc{x3d} pathway introduced by \cite{Vogt2016}. With this approach, the 3-D map of the O-rich ejecta in YSNR 1E\,0102 shown in Fig.~\ref{fig:projections} is made available as an interactive model to the readers of this article via most mainstream web browsers\footnote{For a full list of supported web browsers, see \url{http://www.x3dom.org/contact/} (accessed 2016, September 25).} (and without the need for specific plugins). In its most straightforward implementation, creating an interactive \textsc{html} document only requires some basic \textsc{html} code to load the \textsc{x3d} file (created via \textsc{mayavi} in this case). The \textsc{x3dom} environment however offer several tools that can be easily exploited with additional \textsc{javascript} functions. We explore some of these additional capabilities particularly suited for scientific applications in the next section.

\section{Specific \textsc{x3dom} tools to boost the scientific value of interactive 3-D models}\label{sec:tools}

\cite{Vogt2016} already suggested that \emph{action buttons} allowing the reader to interact with a given 3-D model can significantly enhance the scientific value of an interactive figure. The use of pre-defined viewpoints that can be toggled by the reader allow (for example) to precisely guide their attention to elements of interest, without restricting their ability to explore the model as they please. Action buttons altering the visibility of different elements inside the model can also allow readers to \emph{remove} the outer layers of complex structures to reveal their inner workings.

For the interactive 3-D model of YSNR 1E\,0102, we have implemented additional interaction buttons, with the intention to increase the scientific usefulness of the interactive figure. In addition to the \emph{Viewpoints} and \emph{Toggle} buttons already used by \cite{Vogt2016}, we introduce:
\begin{enumerate}
\item a \emph{Screenshot} ability, which allows readers to easily save a \textsc{png} image of the interactive window in its current state,
\item the display (and live update) of the flux density level corresponding to the outer layer of the [O\,\textsc{\smaller III}]$\lambda$5007 emission visible at any given time, and
\item \emph{clip planes}, which allow slicing of the model in any direction and at any angle. In particular, the offset of the clip plane from the model center (in pc) and their angle of rotation (in degrees, set by the reader) are also updated live.
\end{enumerate} 

The full \textsc{html} and \textsc{javascript} files used to create the online interactive model for YSNR 1E\,0102 have been uploaded onto a new sub-directory within the dedicated Github repository of the \textsc{x3d} pathway created by \cite{Vogt2016}. These files are heavily commented and once again intended as stepping stone for members of the community interested in exploiting the capabilities of the \textsc{x3d} pathway. 

\begin{figure}[htb!]
\centerline{\includegraphics[width=\hsize]{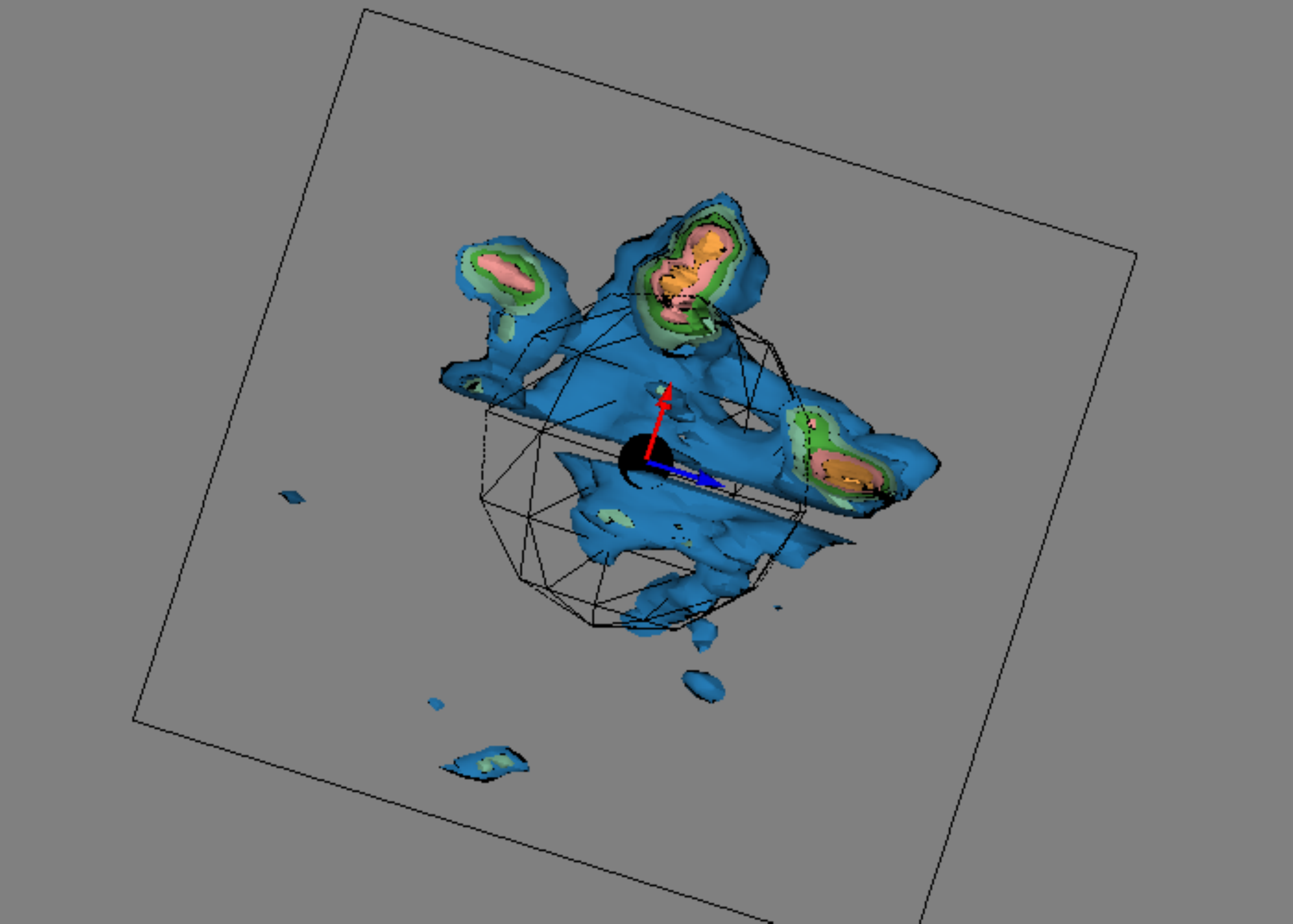}}
\caption{Screenshot of the interactive 3-D map of the oxygen-rich ejecta in YSNR 1E0102, demonstrating some of the associated \textsc{x3dom} visualization \& analysis tools. These include the ability to peel intensity layers (the dark blue layer corresponds to a flux density level of $7\times10^{-17}$ erg\,s$^{-1}$\,cm$^{-2}$\,\AA$^{-1}$), and clip planes to \emph{slice} the model at a certain location and angle. The corresponding state of the interaction buttons is shown in Fig.~\ref{fig:buttons}. An interactive version of this Figure is accessible in the online version of this article. \textit{Note on the arXiv version: until the publication of the article, the online model can be accessed at \url{http://fpavogt.github.io/x3d-pathway/YSNR.html}}.}\label{fig:screenshot}
\end{figure}

\begin{figure*}[htb!]
\centerline{\includegraphics[width=0.9\textwidth]{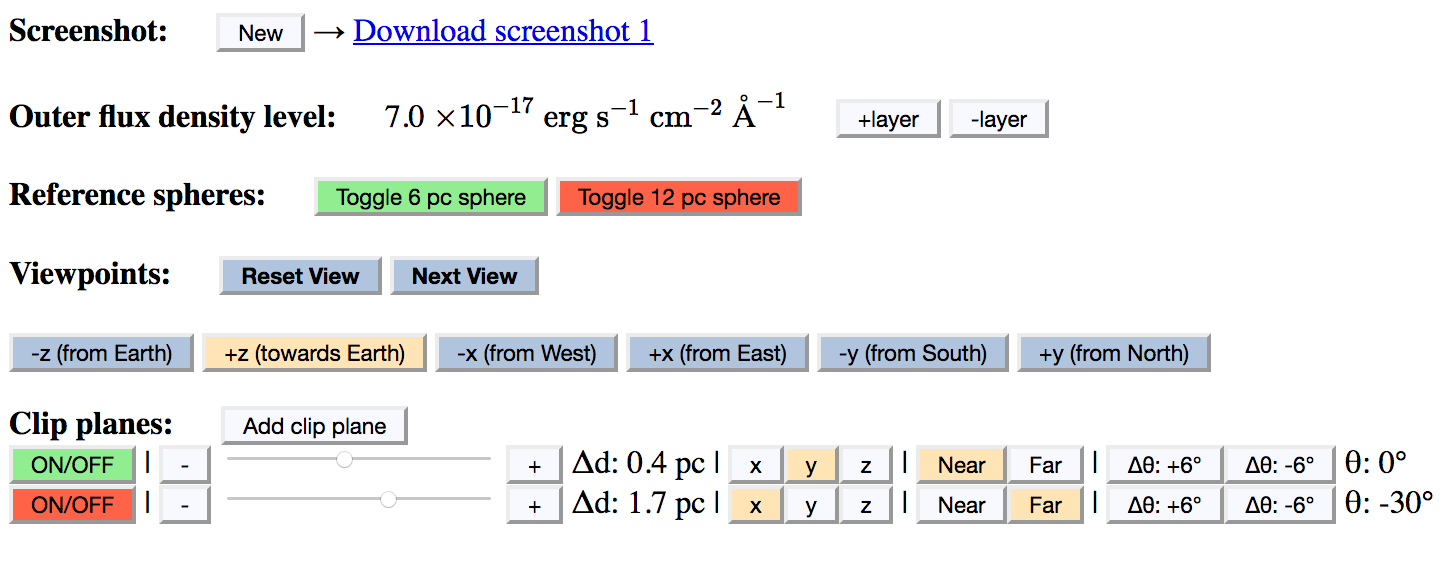}}
\caption{Screenshot of the interaction buttons present on the demonstration \textsc{html} page hosting the 3-D map of the O-rich ejecta of YSNR 1E\,0102. These buttons complement the (basic) \textsc{x3dom} ability to manually slide, zoom-in, zoom-out and rotate around the model; they allow to save screenshots, peel the intensity layers of the model, toggle size-reference spheres on \& off, move to pre-defined viewpoints \& reset the current one, as well as add clip planes to slice the model. Clip planes can be moved in full 3-D space, with their distance from the model center $\Delta$d and rotation angle $\theta$ updated automatically. The  value of the flux density level of the faintest layer visible is also updated automatically. }\label{fig:buttons}
\end{figure*}

The \textsc{javascript} code responsible for the handling of clip planes in our interactive model of YSNR 1E\,0102 has been built upon the demonstration example advertised on the official \textsc{x3dom} website\footnote{\url{http://www.x3dom.org/experimental-clipplane-implementation-available/}} on 2014, July 10. In comparison with the original \textsc{javascript} functions created by ``Timo'' on 2014, June 16, our version of the code contains new features, including the ability to fetch a clip plane location (its translation and rotation angle), the possibility for the user to slide a clip plane \emph{one-step-at-a-time} using dedicated buttons, and the option to set a given clip-plane fully invisible (or not). Most importantly, our version of the code corrects a bug in the original version, in which the handling of a simultaneous translation and rotation led to an erroneous behavior of the outer rectangular frame associated with a given clip plane. A screenshot of the interactive model of YSNR 1E\,0102 demonstrating the use of a clip plane is presented in Fig.~\ref{fig:screenshot}. The corresponding state of the action buttons is shown in Fig.~\ref{fig:buttons}, for completeness. We note that beside the color-changing behavior of the action buttons, our \textsc{html} \& \textsc{javascript} codes include very little \emph{styling} (\textsc{css}-like) elements on purpose, to keep them as simple, clear and concise as possible.

The use of clip plane(s) significantly expands the scientific potential of a given interactive 3-D model. Beyond their primary purpose of \emph{slicing} through the data, they also allow the user to measure distances and angles in 3-D (provided these characteristics are communicated to the user): an important feature that effectively provides to an interactive 3-D diagram a similar level of \emph{exploitability} than that of a more usual 2-D diagram (from which the value of data points can be interpolated). Clip planes are one out of many features accessible through the \textsc{x3dom} framework, but most certainly one of special importance for any scientific applications. We note that the \emph{interactivity} offered by clip planes (and the \textsc{x3dom} framework in general) represents one clear step towards the suggestions of \cite{Goodman2012} that described the importance of interactivity of 3-D diagrams (coupled with the ability to \emph{link} different views and diagrams with one another). 

\section{Summary}\label{sec:summary}

In this article, we have demonstrated the potential of the \textsc{x3d} pathway for optical IFS observations with YSNR 1E\,0102 in the SMC as an example. Using the 3-D map of the oxygen-rich ejecta in this system, we have exploited the \textsc{x3d} pathway introduced by \cite{Vogt2016} to create an interactive 3-D model of YSNR 1E\,0102 hosted in an \textsc{html} document, accessible via most mainstream web browsers. We have explored additional features of the \textsc{x3dom} framework to increase the scientific potential of interactive 3-D diagrams, in particular the possibility of using clip planes to slice through the data. Our corrected and enhanced set of \textsc{javascript} functions designed to handle clip planes (originally shared on the \textsc{x3dom} website), and allowing users to extract scientifically valid information from a given 3-D model -- i.e. distances and angles -- is made freely available to the scientific community on a dedicated Github repository.

At the time of publication of this article, the \textsc{x3d} pathway is already actively supported by leading journals in the field of astrophysics. The  generation of 3-D models (and their export to the \textsc{x3d} format) thus most certainly remains the biggest hurdle in terms of implementation and a clear obstacle slowing the expansion of the \textsc{x3d} pathway in the field of astrophysics. The \textsc{mayavi} module in \textsc{python} offers a simple, \emph{pythonic} tool for astronomers to create 3-D models, although specific limitations still impede on the creation of more advanced products \citep{Punzo2015,Vogt2016}. \textsc{blender} is an open-source alternative to \textsc{mayavi} that offers a nearly unlimited set of options and tools, but at the price of an extremely steep learning curve. From that perspective, recent efforts like \textsc{frelled} \citep{Taylor2015} and \textsc{astroblend} \citep{Naiman2016} aiming at bringing \textsc{blender} closer to astronomers are certainly worth mentioning \citep[see also][]{Kent2013,Kent2015}. 

The \textsc{x3dom} framework is oblivious to the manner a given 3-D model is generated. It is therefore of no importance (from the perspective of the \textsc{x3d} pathway) whether the \textsc{mayavi} limitations will be gradually addressed or not, whether tools such as \textsc{astroblend} and \textsc{frelled} will become increasingly popular within our community, or whether other dedicated software solutions will keep emerging \citep[see e.g.][]{Barnes2006,Steffen2011,Woodring2011,Punzo2016}: that is, so long as our tools support model exports to the \textsc{x3d} format. 

Most importantly, the future of the \textsc{x3dom} framework, intended to be the way that 3-D models are shared on the World Wide Web, is tied to much more than its sole use within the field of astrophysics. From that perspective, we believe that our choice as a community does not lie in deciding on the \emph{likelihood of a future} for this technology, but rather in how soon we shall actively embrace the potential of the \textsc{x3d} pathway to improve our ability to share \& publish complex multi-dimensional datasets. We note that the creation of a dedicated \emph{astrophysics working group} within the Web3D Consortium\footnote{\url{http://www.web3d.org}} would give our community the ability to exert a direct influence on the evolution of the \textsc{x3dom} framework, in a similar fashion to that already pursued by the \emph{medical working group} towards a better support of human anatomy representation.

\acknowledgments
We thank the anonymous referee for a prompt and constructive report. This research has made use of the following \textsc{python} packages: \textsc{statsmodel} \citep{Seabold2010}, \textsc{matplotlib} \citep{Hunter2007}, 
\textsc{astropy}, a community-developed core Python package for Astronomy \citep{AstropyCollaboration2013}, \textsc{aplpy}, an open-source plotting package for
\textsc{python} hosted at \url{http://aplpy.github.com} and \textsc{mayavi} \citep[][]{Ramachandran2011}. This research has also made use of \textsc{montage}, funded by the National Science Foundation under Grant Number ACI-1440620 and previously funded by the National Aeronautics and Space Administration's Earth Science Technology Office, Computation Technologies Project, under Cooperative Agreement Number NCC5-626 between NASA and the California Institute of Technology, of the \textsc{aladin} interactive sky atlas \citep{Bonnarel2000}, of \textsc{saoimage ds9} \citep{Joye2003} developed by Smithsonian Astrophysical Observatory, and of NASA's Astrophysics Data System.
 
FPAV ESO's fellowship is co-funded under the Marie Curie Actions of the European Commission (FP7-COFUND). IRS was supported by Australian Research Council Laureate Grant FL0992131. AJR is thankful for funding provided by the Australian Research Council Centre of Excellence for All-sky Astrophysics (CAASTRO) through project number CE110001020.

{\it Facilities:} \facility{ATT}.

\bibliographystyle{astroads} 
\bibliography{bibliography_fixed} 

\end{document}